\documentclass[twocolumn,aps,pra,showpacs,raggedbottom,nobalancelastpage,amssymb,superscriptaddress]{revtex4-1}

\usepackage{epsfig,amssymb,amsmath,latexsym}
\usepackage{subfigure}
\usepackage{wrapfig}
\usepackage{float}
\usepackage{dsfont}
\newcommand\beq{\begin{equation}}
\newcommand\eeq{\end{equation}}

\newcommand\bea{\begin{eqnarray}}
\newcommand\eea{\end{eqnarray}}

\newcommand\bi{\begin{itemize}}
\newcommand\ei{\end{itemize}}

\newcommand\wvsd{{\textsf{WVs}}}

\newcommand\wvd{{\textsf{WV}}}
\newcommand\wv{{\textsf{WV~}}}
\newcommand\ctapd{{\textsf{CTAP}}}
\newcommand\ctap{{\textsf{CTAP~}}}
\newcommand\qpcd{{\textsf{QPC}}}
\newcommand\qpc{{\textsf{QPC~}}}

\newcommand{\tdel}{t_{\rm delay}}
\newcommand{\tmax}{t_{\rm max}}
\newcommand{\omax}{\Omega_{\rm max}}

\def\ra{\rangle}

\newcommand{\ave}[1]{\big\langle #1\big\rangle}
\newcommand{\weak}[3]{{\phantom{\big\rangle}}_{#1}\ave{#2}_{#3}}
\newcommand{\braket}[2]{\left<\left.{#1}\right|{#2}\right>}

\newcommand{\ket}[1]{\left|{#1}\right>}
\newcommand{\bra}[1]{\left<{#1}\right|}

\newif\ifboo \boofalse

\begin{document}
\title{Sensing electrons during an adiabatic coherent transport passage}
\author{Oded Zilberberg}
\affiliation{Institute for Theoretical Physics, ETH Zurich, 8093 Z{\"u}rich, Switzerland}
\author{Alessandro Romito}
\affiliation{Department of Physics, Lancaster University, Lancaster LA1 4YB, United Kingdom}

\date{\today}
\begin{abstract}
We study the detection of electrons undergoing coherent
transfer via adiabatic passage (\ctapd) in a triple quantum-dot system with a quantum point-contact sensing the change of the middle dot.
In the ideal scenario, the protocol amounts to perfect change transfer between the external dots with vanishing occupation of the central dot at all times, rendering the measurement and its backaction moot. Nevertheless, even with minor corrections to the protocol, a small population builds up in the central dot. We study the measurement backaction by a Bayesian formalism simulation of an instantaneous detection at the time of maximal occupancy of the dot. 
We show that the interplay between the measurement backaction and the non-adiabatic dynamics induce a change of the success probability of the protocol, which quantitatively agrees with a continuous detection treatment. 
We introduce a correlated measurement signal to certify the non-occupancy of the central dot for a successful \ctap protocol, which, in the weak measurement limit, confirms a vanishing occupation of the central dot. Our proposed correlated-signal purports that proper experimental method by which to confirm \ctapd.
\end{abstract}

\maketitle

\section{Introduction} 
Quantum measurements constitute one of the main pillars of quantum mechanics. They induce an unavoidable backaction on the measured system~\cite{von-neumann}. 
This trait can be advantageously used for applications in quantum information processing, ranging from error correction~\cite{Nielsen2000}, to improved quantum state discrimination~\cite{Zilberberg2013}, and to quantum feedback~\cite{Vijay2012}. On the other hand, the impact of measurement backaction can be particularly detrimental in the detection of quantum coherent processes, as the measurement corresponds to a strong decoherence channel~\cite{wiseman2009quantum}. The regime of weak measurement, in which the backaction is reduced alongside the rate of information acquisition, is therefore of particular interest. Weak measurements, in fact, enable detection while minimally disturbing the coherent process and make it possible to define meaningful conditional outcomes in quantum regimes~\cite{Aharonov1988, Dressel2014}.

\begin{figure}[ht!]
\includegraphics[width=\columnwidth]{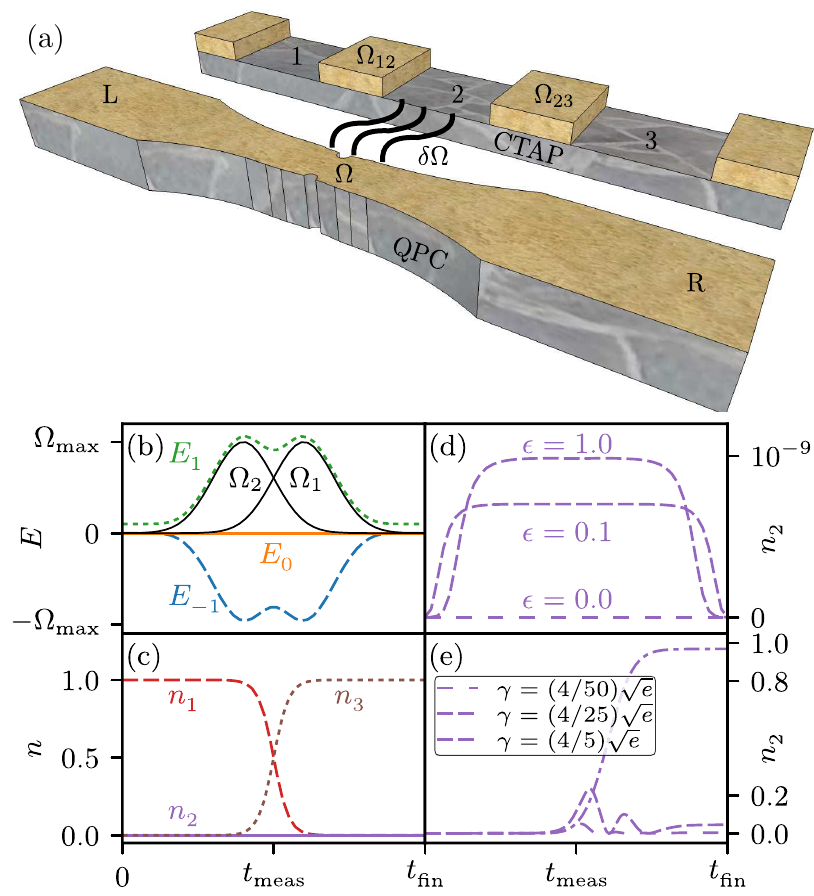}
\caption{\label{Fig:1} The \ctap protocol. (a) Sketch of a  solid state system implementing a \ctap using a triple single-level quantum-dot setup with tunable tunnel barriers. The \ctap transfers the electron from well 1 to well 3 without charge occupancy in the central well 2. A \qpcd detector weakly senses the charge on the central dot. (b) The instantaneous eigenvalues of the \ctap Hamiltonian, [cf.~Eqs.~\eqref{eq-h3w} and \eqref{GauPuls}] for $\epsilon_1=\epsilon_3=0$ and $\epsilon=\epsilon_2=\omax/10$. (c) The time-dependent occupancy of the three wells for $\epsilon_i=0$, with $i=1,2,3$. (d) The time-dependent occupancy of the central well for finite values of the central-well energy $\epsilon=\epsilon_2$, and (e) finite-adiabatic parameter $\gamma$. Both a finite $\epsilon=\epsilon_2$ and a finite $\gamma$ lead to a non-vanishing charge on the central well. The numerical time evolution in (e) is computed with $\delta t=5 \times 10^{-4}\hbar/\omax$.}
\end{figure}

The detection of coherent quantum processes is  relevant for the study of quantum transport. Quantum effects play a crucial role in electronic transport through nanostructures and have been at the core of mesoscopic physics since its foundation. The direct detection of quantum processes by weak measurements is, however, a more recent development. A paradigmatic example thereof involves a which-path detection in  electronic interferometers~\cite{Aleiner1997, OZilberberg2011, Weisz2012}. More recently, the direct detection of electronic transport through virtual state transition in cotunneling processes has been addressed theoretically~\cite{Romito2014,Zilberberg2014}, showing that weak measurements make it possible to collect information on the system through conditional quantities, without destroying the coherent cotunneling process. The adverse effect of backaction on such transport has also been predicted~\cite{Zilberberg2014} and consequently measured~\cite{bischoff2015measurement}.

The role of non-invasive detection of quantum transport processes admits an extra layer of complexity when an external time-dependent driving is applied to the system. A relevant paradigmatic case is that of  coherent transfer via adiabatic passage (CTAP)~\cite{greentree,schoerer,qdot-proposal,Cole2008,Das2009,Menchon2016}.  The \ctap scheme amounts to transporting an electron between two quantum wells (left-to-right) through an additional central well via dynamically tuned tunnel barriers. For appropriate adiabatic driving of the system, the protocol fully transfers the particle, while maintaining a vanishing population at the central well at any time. Thus, the \ctap scheme is manifestly robust against fluctuations that couple to the charge of the central island. Additionally, it is an all-electrical spatial implementation of a well-known quantum optics techniques to transfer populations between long-lived atomic-levels~\cite{Vitanov2001}. There are various proposals to realize the \ctap in different physical systems~\cite{Menchon2016}, and a classical analog of the scheme has been experimentally realized in optics~\cite{photons} with follow-up applications \cite{Menchon2016}.  

The detection of a vanishing charge in the central well along with a successful left-to-right transfer is a striking signature of the \ctap mechanism. At the same time, however, the detector backaction affects the quantum interference underlying the adiabatic passage.
Indeed, a strong projective measurement would destroy the coherence of the central well and correspondingly disrupt the adiabatic passage. Nevertheless, the central well occupation can be addressed by continuous weak measurements. In a recent work~\cite{Rech2011}, the probability distribution function of the current signal of a quantum point contact (\qpcd) sensing the charge in the central dot during a \emph{single-shot} \ctap, as well as, the fidelity of the transport were numerically computed. The gradual acquisition of information on the system was shown to induce loss of fidelity to the population transfer, namely, it appears that the combined detection of the central-well occupation alongside a successful adiabatic passage is unattainable. 

In this work, we analytically study the detection of the charge in the central dot in a \ctap scheme conditional to a successful electron transfer. This quantity provides a direct evidence of the vanishing population of the central dot during a successful \ctap. We show that the detection process is effective in a limited time window at the maximal occupancy of the central well, thus enabling us to introduce an efficient description of the probability distribution function of the detector's signal. Our approach allows us to determine the measured occupation of the central well conditional to the electron passage in the form of so-called weak values (\wvd)~\cite{Aharonov1988}. Our results confirm a vanishing central-well occupation in the limit of weak-measurement backaction and adiabatic evolution.  Interestingly, the \wvd of the central-well population, conditional on an unsuccessful \ctap transfer, approaches a finite negative value, thus providing an indirect evidence of the quantum coherence of the process. Such correlated detection can prove valuable in sensing of other types of prominent adiabatic passage processes,  e.g., in topological pumps~\cite{Kraus2012, Verbin2015, Lohse2016, Zilberberg2018, Lohse2018, Petrides2018, Tambasco2018}.

\section{The \ctap scheme} 
We consider a system consisting of three single-level quantum dots/wells with energy levels $\epsilon_i$ coupled to a quantum point contact (QPC) which serves as charge detector of the occupancy of the central dot/well, see \ref{Fig:1}(a). The external wells, $1$ and $3$, are connected to the central one, $2$, by time-dependent tunneling rates $\Omega_{12} (t)$, $\Omega_{23} (t)$. The Hamiltonian of the system is written as
\begin{equation}
H_{3w} = \sum_{i=1}^3 \epsilon_i c_i^{\dagger} c_i + \left( \hbar\Omega_{12}(t) c_1^{\dagger} c_2
+ \hbar\Omega_{23}(t) c_2^{\dagger} c_3 + \rm{h.c.} \right) ,
\label{eq-h3w}
\end{equation}
where $c_i^{\dagger}$ creates an electron in well $i$.
Provided that the energy levels of the external wells are the same , $\epsilon_1=\epsilon_3=0$, the \ctap protocol coherently transfers an electron from well $1$ to well $3$ by applying Gaussian voltage pulses to tune the tunnel barriers in time~\cite{greentree}
\begin{eqnarray}
\Omega_{12}(t) & = & \omax \exp \left[ - \frac{(t - \tmax/2 - \tdel)^2}{2 \sigma^2} \right] \nonumber \\
\Omega_{23}(t) & = & \omax \exp \left[ - \frac{(t - \tmax/2)^2}{2 \sigma^2} \right] ,
\label{GauPuls}
\end{eqnarray}
where both pulses have the same height, $\omax$, and width, $\sigma$, and are delayed by $\tdel$. The probability of transferring the electron is maximal when $\sigma =\tmax / 8$ and $\tdel = 2 \sigma$~\cite{Rech2011}, and in the ideal adiabatic limit, it approaches $1$.

The deterministic success of the \ctap relies on the Hamiltonian (\ref{eq-h3w}) with $\epsilon_1=\epsilon_3=0$ having a zero-energy, $E^{\phantom{\dagger}}_0=0$, eigenstate at any time, as shown in Fig.~\ref{Fig:2} (b). The basic idea is that the time-dependence in Eq.~\eqref{GauPuls} adiabatically evolves the left-well occupancy to the right-well occupancy through that zero-energy eigenstate. Consider for simplicity the case where all $\epsilon_i=0$. At the onset of the protocol, $t_{\text{start}} \to -\infty$, the system's eigenstates are degenerate at zero energy. The switching on of $\Omega_2$ (note that, the coupling $\Omega_2$ between wells $2$ and $3$ is switched on before the coupling $\Omega_1$ between $1$ and $2$) maintains only the left-well state at zero energy. This zero-energy state adiabatically evolves to the right-well state at the end of the protocol, $t_{\text{fin}} \to \infty$. Note, that having $\epsilon_2\neq 0$ does not affect the properties of the zero-energy eigenstate, see Fig.~\ref{Fig:1} (b). 

Ideally, the \ctap process takes infinite time and yields unitary transfer probability, cf.~Fig.~\ref{Fig:1}(c). Realistically, a finite duration, $t_{\text{fin}}-t_{\text{start}}$ introduces a non-zero overlap of the initial left-well state at time $t_{\text{start}}=0$ with the finite-energy eigenstates. Yet, the protocol is designed to maintain a maximal overlap of the remaining zero-energy eigenstate with the left-well state. Note that the success of the protocol is not altered by a finite $\epsilon_2$, since the zero-energy eigenstate is preserved [cf.~Fig.~\ref{Fig:1}(b)] and its initial and final overlaps with the left and right wells are unaffected. 

The adiabaticity of the process is controlled by a generalized Landauer-Zener parameter $\gamma=\max\left|\bra{\psi_1}\partial_t\hat{H}_{3w}\ket{\psi_0}/(E^{\phantom{\dagger}}_1-E^{\phantom{\dagger}}_0)^2\right|= 4 \sqrt{e}/(t_{\textrm{max}}\Omega_{\textrm{max}}) \ll1$, with $E^{\phantom{\dagger}}_j$ and $\ket{\psi_j}$ ($j=-1,0,1$) the instantaneous eigenenergies and eigenstates of $\hat{H}_{3w}$ at time $t$. 
Remarkably, in the adiabatic limit and $\epsilon_2=0$, the occupation of the central well is identically zero \cite{greentree,Rech2011}, $\langle c_2^{\dag}c_2 \rangle \equiv n^{\phantom{\dagger}}_2 =0$, as shown in Fig. \ref{Fig:1}(c). This makes the system insensitive to any external interaction with the central-well population, being it by undesired fluctuations or by a charge detector. 
The features $n^{\phantom{\dagger}}_2(t)=0$ is modified by either $\epsilon_2 \neq 0$ (alongside a finite duration of the experiment) or by diabatic corrections at $\gamma \ne 0$. Hence, a detection process of a \ctap should be considered along with the $\gamma \to 0$ and $\epsilon_2 \to 0$ limits. 
 
The effect of finite $\epsilon=\epsilon_2$ can be accounted for analytically in the adiabatic limit yielding
\begin{eqnarray}
 n^{\phantom{\dagger}}_2(t) &=& \sum_{j,k=-1,0,1} \left(\alpha_{j}^{t_{\text{start}}}\right)^{\star} \bra{\psi_{j}(t)} c_2^{\dag}c_2   \ket{\psi_{j}(t)} \alpha_{j}^{t_{\text{start}}} \nonumber \\ &=& \left\vert \alpha_{1}^{t_{\text{start}}}  \sqrt{\sqrt{4 \Omega_1^2+4\Omega_2^2+\epsilon^2}+\epsilon} \nonumber \right.\\
 & & \left. -\alpha_{-1}^{t_{\text{start}}}  \sqrt{\sqrt{4 \Omega_1^2+4\Omega_2^2+\epsilon^2}-\epsilon}\right\vert^2/2,
\end{eqnarray}
 where $\alpha_{j}^{t_{\text{start}}}\equiv \braket{\psi_{j}(t_{\text{start}})}{1}$ is the overlap amplitude of the left well with the eigenstates at time $t=t_{\text{start}}$. The resulting time-dependent occupation of the central well is reported in Fig. \ref{Fig:1}(d), where a finite, yet small, occupancy $n^{\phantom{\dagger}}_2(t)$ is maintained around $t_{\rm meas}=(\tmax+\tdel)/2$.  

In the finite $\gamma$ case, the evolution of the initial state can be determined numerically.  It can be obtained by discretizing the time in intervals $\delta t$ where the Hamiltonian is assumed to stay constant. We use the Crank--Nicolson method~\cite{Crank:1947,Crank:1996} to approximate the propagator over a time period $\Delta t$. The time evolution of the system is then expressed as 
\begin{align}
\rho(t+\Delta t)=\hat{U}(\Delta t)\rho(t)\hat{U}^\dagger(\Delta t)\, ,
\end{align}
where the propagator in Cayley form~\cite{Goldberg:1967} is
\begin{align}
\hat{U}(\Delta t)=(\mathds{1}+i\frac{\Delta t}{2}\hat{H}_{3w}(t))^{-1}(\mathds{1}-i\frac{\Delta t}{2}\hat{H}_{3w}(t))\, .
\end{align}
The time-evolution is applied to an initial state density matrix elements $\rho$ written in the $\{ |1\ra, |2\ra\, |3\ra\}$ basis at time $t_{\text{start}}=0$.
Our numerical calculations for $\langle c_2^{\dag}c_2 \rangle=n^{\phantom{\dagger}}_2(t)$ reported in Fig. \ref{Fig:1}(e) shows that, for relatively adiabatic evolution, the largest correction to the central-well population occurs at a short time-window around the middle of the pumping protocol, $(\tmax+\tdel)/2$. At other times, $n^{\phantom{\dagger}}_2$ is exponentially small~\cite{Rech2011}.
This makes the protocol exponentially insensitive to external fluctuations on the dot, but, at the same, time poses a limit to the direct detection of the charge in the central well~\cite{Rech2011}. 

\section{The detection process} 
 To determine the effect of the measurement process, we assume that the detector is an ideal quantum point contact (\qpcd) \cite{gurvitz,Korotkov}, whose current is solely sensitive to the presence of an electron in the middle well.
Beside being routinely used in experiments as a charge sensor~\cite{Field1993}, a \qpc provides a simple, yet general, model for a detector. 
The \qpc is characterized by the tunneling amplitudes, $\Omega$, and $\Omega + \delta \Omega$, depending on whether well $2$ is unoccupied or occupied, respectively~\cite{gurvitz}. 
The coupling between the system and the detector is then given by the Hamiltonian
\begin{eqnarray}
H_{qpc} & = & \sum_{r} (E_{r}-\mu_r) a_{r}^{\dagger} a_{r}  + \sum_{l} (E_{l}-\mu_l)  a_{l}^{\dagger} a_{l}   \nonumber \\
& & + \sum_{l,r} \hbar \left( \Omega + \delta \Omega~ c_2^\dagger c_2\right) ( a_{r}^{\dagger} a_{l}
+  a_{l}^{\dagger} a_{r}) ,
\label{H_qpc}
\end{eqnarray}
where $a_{r}^{\dagger}$ and $ a_{l}^{\dagger}$ are the electron creation operators in the right and left electrode
respectively, while $E_{r(l)}$ stands for the set of energy levels in the reservoirs kept at chemical potentials $\mu_{r(l)}$ so that the difference is set by the applied voltage bias $\mu_r-\mu_l=eV$. Here, we assume all tunneling amplitudes to be real and independent of the states in the \qpc leads. We further restrict ourselves to the zero temperature limit, so that no extra noise sources are present and the detector is quantum limited~\cite{Averin2005}. 
The macroscopic (classical) signal in the detector is the current through the \qpcd. This is a stochastic signal whose distribution generically depends on the system's state and the duration of the measurement, $\tau_V$. When the central well is empty the average current is $I_e=e T eV/\hbar$ where $T=2\pi \nu_l \nu_r \Omega^2$ is the transmission probability through the \qpc and $\nu_{r,l}$ the density of states in the leads. Similarly, when the central well is occupied, we have the average current $I_o= e (T+\delta T) eV/\hbar$. 

For a generic (coherent) state of the system, the stochastic current outcome $I$ can be regarded as the fraction of successfully transmitted electrons across the \qpcd, $I=e n/N$  where $N$ is the total number of impinging electrons at a rate $eV /\hbar$, during the measurement time $\tau_V$. The \qpc current is therefore characterized by the probability distribution $P(I,N)$.
For the empty case, the transmission probability is $T$, and for large $N$, $I$ will be normal distributed with variance $4S_I/\tau_V$, where $S_I =2 e I_e (1-T)$ is the current shot-noise. The variance of the distribution is the same for the occupied configuration as long as $\delta \Omega \ll \Omega$. By increasing $N$, the variance is gradually reduced and the state of the well being occupied or unoccupied is resolved. The two states are distinguishable when $ 4S_I/\tau_V < (I_e-I_o)^2$, which is $\tau_V >\tau_M=4S_I/ (I_e-I_o)^2$, where $\tau_M$ is referred as the measurement time. As long as $\tau_V \ll \tau_M$, the measurement is not sufficiently long to distinguish the two states, and we are in the weak measurement regime.

It is convenient to rescale the stochastic current variable $I$ to a dimensionless outcome variable, $x=(2I-I_e-I_o)/(I_o-I_e)$, so that $I_e$ and $I_o$ are linearly mapped to the dimensionless outcome values $-1$ and $1$, respectively. For a given state of the system defined by the density matrix $\rho (t) = \sum_{i,j=1,2,3}\rho_{i,j}\vert i \rangle \langle j\vert$,  the probability distribution of the \qpc current is given by~\cite{Korotkov,jordan}
\begin{equation}
P(x,N)= \left(\rho_{11}+\rho_{33}\right)P(x,N|-1)+\rho_{22}P(x,N|1),
\label{prob dist}
\end{equation}
where $P(x,N|s)$ is a Gaussian distribution with an average of $s$, and a variance of $D/N$, where $D=\tau_M eV/\hbar$, and $D/N \gg 1$ sets the weak measurement limit.  
The measurement backaction alters the state of the system. For a given measurement outcome $x$, the density matrix after the measurement is~\cite{Korotkov,jordan}  
\begin{eqnarray}
\rho^\prime(t+\tau_V,x)  =  \frac{1}{\mathcal{P}} 
\left(
\begin{array}{ccc}
\rho_{11}(t)e^{\alpha} & \rho_{12}(t) & \rho_{13}(t)e^{\alpha} \\
\rho_{12}(t) & \rho_{22}(t)e^{-\alpha} & \rho_{23}(t) \\
\rho_{13}(t)e^{\alpha} & \rho_{23}(t) & \rho_{33}(t)e^{\alpha} \\
\end{array}
\right), 
\label{weak_meas}
\end{eqnarray}
where $\mathcal{P}=\rho_{11}(t)e^{\alpha}+\rho_{22}(t)e^{-\alpha}+\rho_{33}(t)e^{\alpha}$,  and we have introduced $\alpha = {x} N/D$. 
Controlling the duration of the measurement, $\tau_V \to 0$, the Bayesian formalism makes it possible to follow the quantum evolution of the system state during the measurement process~\cite{Korotkov,jordan}.


\begin{figure}
	\includegraphics[width=\columnwidth]{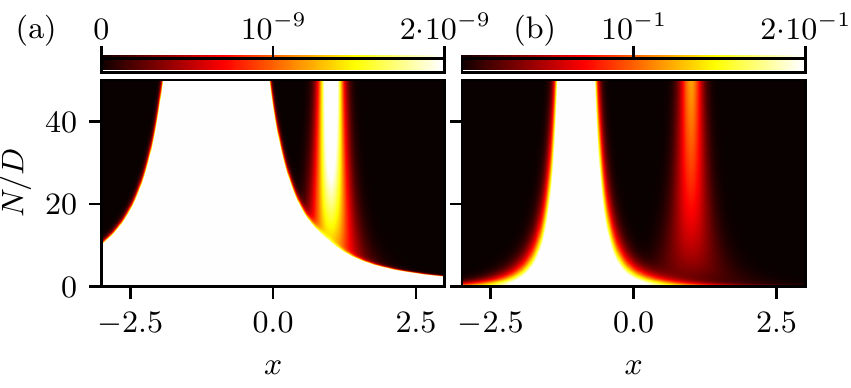}
	\caption[]{\label{Fig:2}
		Detector signal. Probability distribution density of the dimensionless detector signal, $x$ for different measurement strength, $N/D$ at $t_{\rm meas}=(t_{max}+t_{delay})/2$ for (a) $\epsilon/\omax =1/50$ and $\gamma \to 0$ and (b) $\epsilon/\omax =1/50$ and $\gamma/4\sqrt{e}=1/50$. In all simulations for (b) $\delta t=5 10^-2 \hbar/\omax$. The results reproduce with good accuracy those from the full simulation in Ref.~\onlinecite{Rech2011}. The different color scales highlight the acute difference in the to-be-measured accumulated charge in the central well by the two scenarios, see also Figs.~\ref{Fig:1}(d) and (e). }
\end{figure}

Computing the signal of the \qpc and its effect on the efficiency of the \ctap process requires a numerical simulation of the system-detector evolution over the whole cycle, as in~Ref.~\onlinecite{Rech2011}. Taking advantage of the vanishing occupancy of the central well during the \ctap protocol, we can make the numerical computation considerably easier. We first note that in order to probe the vanishing occupancy of the central well, it is sufficient to sense it at the most volatile instance of time when a nonvanishing population can develop, as opposed to following the charge throughout the protocol with negligible chance of detection.  The specific dynamics of the \ctap scheme makes the sensing at that given time as informative as the full charge tracking.
In fact, as shown in Fig~\ref{Fig:1}(e), the population in the central dot, $n^{\phantom{\dagger}}_2$,  becomes appreciable only  around $t_M=(\tmax+\tdel)/2$, before decreasing once more. We therefore expect that a short pulse measurement, a measurement kick, at a single time, when the central-well population is in its maximum, $t_M \approx(\tmax+\tdel)/2$, plays the same role as an integrated charge detection, and that the two descriptions of the system-detector dynamics should essentially capture the same physics. In other words, our simplification decouples the system's numerical time-evolution from that of the detector until the measurement time, $t_{\rm meas}=(\tmax+\tdel)/2$. At that time, the pulsed weak-measurement can be treated analytically. 

We plot $P(x,N)$ in Figs.~\ref{Fig:2}(a) and (b) for the ideal adiabatic limit with $\epsilon\equiv\epsilon_2\neq 0$ and for a finite adiabatic parameter $\gamma$, respectively. The  probability density distribution we obtain in the adiabatic limit agrees extremely well with the one obtained via a conservative numerical ensemble averaging~\cite{Rech2011}. The difference between the two plotted distributions arises due to the profound difference in the to-be-measured accumulated charge in the central well, i.e., the central well potential $\epsilon\neq 0$ generates a much smaller signal than the non-adiabatic correction for the chosen parameters, see also Figs.~\ref{Fig:1}(d) and (e).



\section{Measurement backaction  and conditional signal}

The advantage of the Bayesian approach is the possibility to address the backaction of \emph{any single measurement}, and not only its \emph{average} effect.
We can thus determine the average outcome of sensing the charge on the central well ($2$) conditional to the success of the pumping cycle
\begin{equation}
\weak{w}{x}{1} = \int\limits_{-\infty}^\infty {x} P({x} |
w) d x = \int x \frac{P(w|x) P(x)}{ P(w)}  dx.
\label{wv integral}
\end{equation}
The expression involves the probabilities of finding, at time $t_{\text{end}}=\tmax+\tdel$, the pumped electron at a given well ($w\in \{1,2,3\}$) given a specific measurement outcome $x$, i.e., $P(w|x )=\langle w | U \rho'(t_{\rm meas},x) U^{-1}| w \rangle$ with $U$ the time-propagator from $t_{\rm start}$ to $t_{\rm meas}$, and the probability of finding the particle in the well $w$, $P(w)=\int\limits_{-\infty}^{\infty} P(w |x) dx$. With the introduced rescaling of the detector signal, the conditional detector outcome is directly translated to the conditional occupancy of the dot via $n^{\phantom{\dagger}}_2=(x+1)/2$ and equivalently $\weak{w}{n^{\phantom{\dagger}}_2}{1}=\left(\weak{w}{x}{1}+1\right)/2$.
Note that, in our numerical method, the time evolution in the calculation of $P(w|x )$ can be conveniently absorbed in the back-in-time evolution of the state $\vert w \rangle$ rather than in the evolution of the density matrix.

\begin{figure}
	\includegraphics[width=\columnwidth]{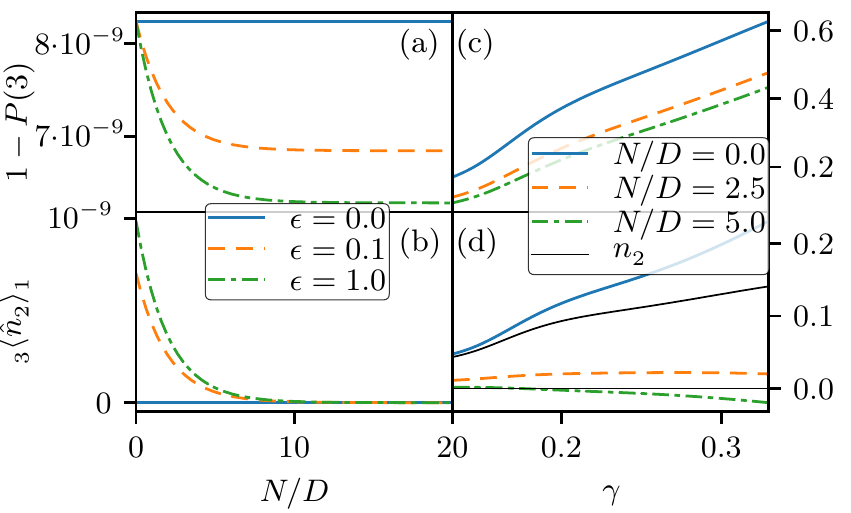}
	\caption[]{\label{Fig:3} Charge measurement for successful \ctapd. Probability of not-finding the electron in the final (right) well at the end of the protocol (a) and corresponding conditional occupancy of the central well, $\weak{2}{n_2}{1}$ (b),  at $t_{\rm meas}=(t_{max}+t_{delay})/2$ as a function of the measurement strength in the ideal adiabatic limit. Panels (c) and (d) plot the dependence of the same variables as in (a) and (b), respectively as a function of the adiabatic parameter, $\gamma$. The success probability $P(3)$ increases with the measurement strength which statistically suppresses unwanted components of the system state in the central well. All results are obtained with $\delta t= 5 \times 10^-2 \hbar/\omax$ and $t_max=50 \hbar/\omax$.}
\end{figure}

In the limit of weak measurement, $D/N \gg 1$, the conditional detector outcome in Eq.~(\ref{wv integral}) takes the form of so-called weak values of the population of the middle well. Weak values  (\wvsd) were introduced as the distinctive result~\cite{Aharonov1988} of measurements consisting of 
(i) initializing the \emph{system} in a certain state $\ket{\psi_i}$---\emph{preselection},
(ii) weakly measuring an observable $\hat{A}$ of the system via a von
Neumann interaction~\cite{von-neumann} 
with a detector, and (iii) retaining the detector output only if the system
is eventually measured to be in a chosen final state, $\ket{\psi_f}$---\emph{postselection}.
The average signal of the detector will then be proportional
to the real (or, possibly, the imaginary) part of the \wv $
\weak{f}{A}{i}=\frac{\bra{\psi_f}\hat{A}\ket{\psi_i}}{\braket{\psi_f}{\psi_i}}$.
Apart from the role of weak values in addressing conceptual questions~\cite{Steinberg1995,Lundeen2011, Aharonov2002} and their use for precision measurements~\cite{Ritchie1990,Pryde2005,Hosten2008,Dixon2009,Starling2009,Brunner:2010,Starling:2010b,jordan,Romito2008,OZilberberg2011}, they provide a way to define conditional physical observables independent of the detector's details~\cite{Choi2013,Romito2014,Zilberberg2014}. In the present case too, the weak value is the proper quantity to address  the detection of the charge in the central well for successful \ctap adiabatic transfers. This coincides with  the conditional signal outcome introduced in Eq.~(\ref{wv integral}). 

The measurement backaction effects are presented in Fig.~\ref{Fig:3}, where the failure probability of \ctap, $1-P(3)$, and the conditional occupancy of the well, $\weak{3}{n^{\phantom{\dagger}}_2}{1}$, are presented. In the ideal adiabatic limit, already in the absence of the measurement, the success probability of the \ctap  and the occupancy of the central well are not optimal, i.e., there exists a finite \ctap failure probability and non-vanishing central well population, see Figs.~\ref{Fig:3}(a) and (b). This is the result of of the finite-time duration of the protocol and the finite energy, $\epsilon$. theoretically expected occupancy of the dot is correctly reproduced by the simulated conditional signal of the detector in Fig.~\ref{Fig:3}(b). Furthermore, the success of the \ctap is surprisingly increasing with increasing measurement strength, and the conditional measured charge on the central well is correspondingly reduced. This effect can be explained by noting that, given the low occupancy of the dot in the absence of measurement, the measurement backaction tends to statistically project the system onto the state with an empty central well. This reduces the unwanted weight of the system state on the central well, thus making the state closer to the ideal \ctap state. Indeed, the more likely such a projection is, e.g., by increasing $\epsilon=\epsilon_2$, the more the measurement backaction can correct for the finite-duration error, see Figs.~\ref{Fig:3}(a) and (b). 

A similar effect of backaction occurs for the case of diabatic corrections, see Figs. \ref{Fig:3}(c) and (d). In the absence of measurement, the failure probability [Fig. \ref{Fig:3}(c)] and the conditional central-well occupation [Fig. \ref{Fig:3}(d)] reduce to those set by the initial finite duration of the protocol. The occupancy increases and the success probability decreases upon increasing the diabatic corrections. The conditional signal of the detector follows the unconditional occupancy of the well in the adiabatic regime, but deviates for large $\gamma$ since the diabatic dynamics considerably changes the success probability. Also in this case, we see that the measurement backaction plays in favor of the \ctap protocol: it reduces the occupancy of the well [Fig. \ref{Fig:3}(d)], and increases the success probability [Fig. \ref{Fig:3}(c)]. The rational is again that the measurement reduced the unwanted state component on the central well. However, this does not hold when the occupancy of the central well starts deviating considerably from zero and the success probability is low. In this regime, the conditional charge deviated considerably from the unconditional one, and the coherence of the quantum evolution shows up in peculiarities of the conditional value, e.g. the negative values of $\weak{3}{n_2}{1}$.

The conditional occupancy in Figs.~\ref{Fig:3}(b) and (d) in the limit of weak measurement and adiabatic dynamics is a direct measurement of the vanishing 
occupation of the central well when restricting to successful \ctap processes. 
This has to be contrasted with an unconditional measurement for which there is no guarantee that the probability of a successful electron transfer, results from \ctap alongside a vanishing central-well occupation.  
Also, as shown in Ref.~\onlinecite{Rech2011}, any measurement asserting an unambiguous value of the central-well population in a single run of a \ctapd, would hinder the success of the protocol, making it inconclusive.

\begin{figure}
\includegraphics[width=\columnwidth]{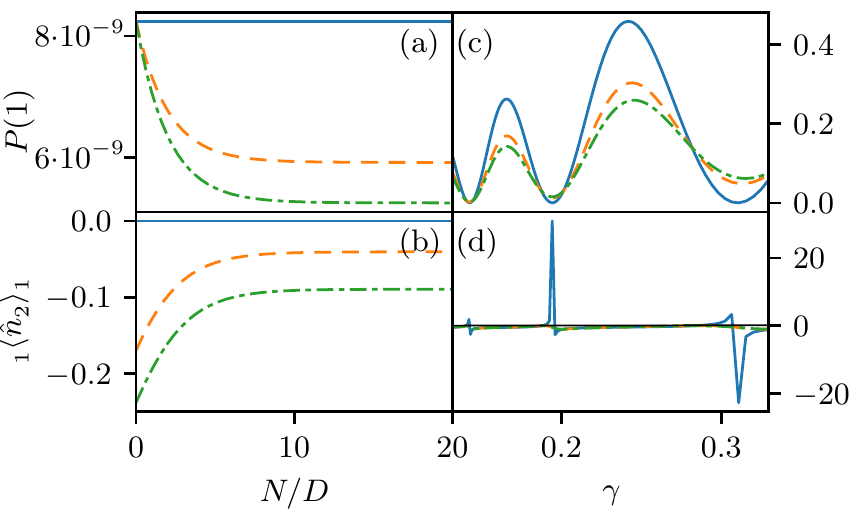}
\caption[]{\label{Fig:4} Unsuccessful \ctapd. Probability of the finding the electron in the initial (left) well at the end of the protocol (a) and corresponding conditional occupancy of the central well (b) at $t_{\rm meas}=(t_{max}+t_{delay})/2$  as a function of the measurement strength in the ideal adiabatic limit. Panels (c) and (d) report the dependence of the same quantities on the adiabatic parameter. The decreasing of $P(1)$ with the measurement strength  is consistent with the results in Fig. \ref{Fig:3}. $P(1)$ has a recurring behaviour as a function of the adiabatic parameter. In correspondence with $P(1) \to 0$, the conditional occupancy of the dot shows large or negative values characteristic of peculiar weak values. All results are obtained with $\delta t= 5 \times 10^-2 \hbar/\omax$ and $t_{max}=50 \hbar/\omax$.}
\end{figure}

Interestingly, we can access, via the Bayesian formalism~\cite{Korotkov,jordan}, the conditional value for unsuccessful \ctapd, given by the probability of finding the electron in the left or central wells. The results for $P(1) $ and $\weak{1}{x}{1}$, are shown in Fig.~\ref{Fig:4}. Analogous results are obtained for $P(2)$ and $\weak{2}{x}{1}$.
We note the decreasing of $P(1)$ in Figs.~\ref{Fig:4}(a) and (c) with the measurement strength, which is consistent with the increasing dependence of $P(3)$ in Fig.~\ref{Fig:3}. The dependence on the adiabatic parameter, $\gamma$, shows a recurring behavior. 
While the probability of failing the \ctapd, $P(1)+P(2)$, is a monotonous function of $\gamma$, cf.~Fig.~\ref{Fig:3}(c), the probabilities of detecting the particle in either of the two wells is determined by the quantum evolution in the finite Hilbert space of the system, which generically shows revival as a function of time or system parameters.
 
As expected, $P(1)$ is small in the adiabatic limit. The corresponding conditional occupation of the central well is negative. This non-classical feature is an indirect signature  of the quantum evolution of the \ctapd. In fact, as known from weak measurement theory~\cite{Aharonov1988}, these peculiar features  are in one-to-one correspondence with the violation of certain Leggett-Garg inequalities~\cite{Williams2008}, which set classical (i.e. from macroscopic realism) inequalities for correlated outcomes of a sequence of measurements. Specifically, if we indicate as $x_j(t)$ the dimensionless signal of a \qpc detector coupled to the $j$-th quantum well at time $t$, the classical constraint  $0 \leqslant \vert \weak{1}{x_1(t-{\rm meas})}{1} \vert \leqslant 1$ on the conditional outcome is violated in the limit of weak measurements if and only if the Leggett-Garg inequality, $-3 \leqslant \mathcal{B} \leqslant 1$ with
$\mathcal{B} \equiv \langle x_1(t_{\rm start}) x_2 (t_{\rm meas}) \rangle + \langle x_2(t_{\rm meas}) x_1 (t_{\rm fin}) - \langle x_1(t_{\rm start}) x_1 (t_{\rm fin})  \rangle$, 
is violated too.
%

\section{Summary and Conclusion}
In the present work, we address the detection of the central-well occupancy in a \ctap along with the corresponding backaction. 
We model the measurement as a an instantaneous process taking place at the time of maximal occupancy of the central well, thus decoupling the measurement from the system evolution.
The instantaneous detection reproduces the results of a continuous detection of the central-well occupancy during the entire pumping protocol and allows us to conveniently define and compute the population of the central well conditional to successful electron transfer via \ctapd. 
This quantity, as opposed to single-shot measurements and unconditional averages, is the one that directly probes the occupation of the central well for the adiabatic transfer.
By analysing the weak-measurement limit, we show that the conditional occupation of the central well vanishes in the adiabatic limit, thus providing a direct measurable evidence of the main feature of \ctapd.
We also find that the occupation conditional to a non-successful pumping remains finite in the adiabatic limit, which provides evidence of the coherent quantum nature of the process. Our work puts forward correlated detection as a valuable method for sensing adiabatic passage processes,  e.g., in topological pumps~\cite{Kraus2012, Verbin2015, Lohse2016, Zilberberg2018, Lohse2018, Petrides2018, Tambasco2018}.

\section{Acknowledgements} We thank A.~Saha and Y.~Gefen for fruitful discussions. OZ acknowledges financial support from the Swiss National Science Foundation. AR acknowledges research support from  EPSRC (Grant EP/P030815/1).

\bibliographystyle{apsrev4-1} 

%
%
%

\end{document}